\documentclass[twocolumn,prl,showpacs]{revtex4}
\usepackage{graphicx}
\begin{document}

\newcommand{\re}{\mathop{\mathrm{Re}}}

\newcommand{\be}{\begin{equation}}
\newcommand{\ee}{\end{equation}}
\newcommand{\bea}{\begin{eqnarray}}
\newcommand{\eea}{\end{eqnarray}}


\title{How far is it to a sudden future singularity of pressure?}

\author{Mariusz P. D\c{a}browski}
\email{mpdabfz@sus.univ.szczecin.pl}
\affiliation{\it Institute of Physics, University of Szczecin, Wielkopolska 15,
          70-451 Szczecin, Poland}
\author{Tomasz Denkiewicz}
\email{atomek@sus.univ.szczecin.pl}
\affiliation{\it Institute of Physics, University of Szczecin, Wielkopolska 15,
          70-451 Szczecin, Poland}
\affiliation{\it Fachbereich Physik, Universit\"at Rostock,
Universit\"atsplatz 3, D-18051 Rostock, Germany}

\author{Martin A. Hendry}
\email{martin@astro.gla.ac.uk}
\affiliation{\it Department of Physics and Astronomy,
University of Glasgow, Glasgow G12 8QQ, UK}

\date{\today}

\input epsf

\begin{abstract}
We discuss the constraints coming from current observations of type
Ia supernovae on cosmological models which allow sudden future
singularities of pressure (with the scale factor and the energy
density regular). We show that such a sudden singularity may happen
in the very near future (e.g. within ten million years) and its
prediction at the present moment of cosmic evolution cannot be
distinguished, with current observational data, from the prediction
given by the standard quintessence scenario of future evolution.
Fortunately, sudden future singularities are characterized by a
momentary peak of infinite tidal forces only; there is no geodesic
incompletness which means that the evolution of the universe may
eventually be continued throughout until another ``more serious''
singularity such as Big-Crunch or Big-Rip.
\end{abstract}

\pacs{98.80.Cq; 98.80.Hw}

\maketitle

Over the past decade observations of high-redshift Type Ia
supernovae (SNIa) have provided strong evidence that the expansion
of the universe is accelerating, driven in the standard paradigm by
some form of dark energy\cite{supernovaenew,superrecent}. Current
data\cite{superrecent} continue to leave open the possibility that
dark energy exists in the form of phantom energy, which may violate
all energy conditions \cite{phantom}: the null ($\varrho c^2 + p
\geq 0$), weak ($\varrho c^2 \geq 0$ and $\varrho c^2 + p \geq 0$),
strong ($\varrho c^2 + p \geq 0$ and $\varrho c^2 + 3p \geq 0$), and
dominant energy ($\varrho c^2 \geq 0$, $-\varrho c^2 \leq p \leq
\varrho c^2$) conditions (where $c$ is the speed of light, $\varrho$
is the mass density in $kg m^{-3}$ and $p$ is the pressure). Phantom
matter may dominate the universe in the future and drive it towards
a Big-Rip (BR) singularity in which all matter will be dissociated
by gravity \cite{caldwellPRL}. This is dramatically different from
the standard picture of future cosmic evolution which suggests an
asymptotically empty de-Sitter state driven by the cosmological
constant or quintessence \cite{dave98} and leading to the violation
of the strong energy condition only.

Phantom-driven scenarios have encouraged the study of other exotic
possibilities for the future evolution of the universe. One of these
possibilities appears in those models which do not assume any
explicit form for the equation of state $p=p(\varrho)$, leaving the
evolution of the energy density and pressure unconstrained. This
freedom may result in a so-called {\em sudden future singularity\/}
(SFS) of pressure \cite{barrow04} which violates only the dominant
energy condition. The nature of a sudden future singularity is
different from that of a standard Big-Bang (BB) singularity, and
also from a Big-Rip singularity, in that it does not exhibit
geodesic incompletness and the cosmic evolution may eventually be
extended beyond it \cite{lazkoz,adam}. The only physical
characteristic of these singularities is a momentarily infinite peak
of the tidal forces in the universe. In more general models this
peak may also appear in the derivatives of the tidal forces. It is
interesting to note that these types of singularity are in a way
similar to yet another type, which were termed finite density
singularities \cite{dabrowski93}. However, the crucial
difference is that finite density singularities occur as
singularities in space rather than in time, which means that even at
the present moment of cosmic evolution they could exist somewhere in
the Universe \cite{inhpress}. We will not discuss in
detail finite density singularities in this paper since they
basically appear in cosmological models without homogeneity. On the
other hand, it is worth mentioning that the sudden future
singularities are quite generic since they may arise in both
homogeneous \cite{barrow042} {\em and \/} inhomogeneous \cite{sfs1}
models of the universe.

In order to obtain a sudden future singularity consider the simple
framework of an Einstein-Friedmann cosmology governed by the
standard field equations
\bea \label{rho} \varrho &=& \frac{3}{8\pi G}
\left(\frac{\dot{a}^2}{a^2} + \frac{kc^2}{a^2}
\right)~,\\
\label{p} p &=& - \frac{c^2}{8\pi G} \left(2 \frac{\ddot{a}}{a} +
\frac{\dot{a}^2}{a^2} + \frac{kc^2}{a^2} \right)~,
\eea
where the energy-momentum conservation law
\be \label{conser} \dot{\varrho} = - 3 \frac{\dot{a}}{a}
\left(\varrho + \frac{p}{c^2} \right)~,
\ee
is trivially fulfilled due to the Bianchi identity. Here $a(t)$ is
the scale factor, $G$ is the gravitational constant, and the
curvature index $k=0, \pm 1$. What is crucial to obtain a sudden
future singularity is that no link between the energy density and
pressure (the equation of state) is specified. This allows us to
integrate (\ref{conser}) only by quadratures as
\bea \label{quadra} &&\varrho a^3 = \exp \left[- \left.
\left(\frac{3p(t')}{c^2\varrho(t')}\ln{a(t')}\right)
\right\vert_{t_0}^{t} \right. \\&& + \left.
\frac{3}{c^2}\int_{t_0}^{t} \left(
\frac{p(t')}{\varrho(t')}\right)^{\cdot} \ln{a(t')} dt' \right]~.
\nonumber \eea
Of course (\ref{quadra}) reduces to the standard expression for
energy conservation, $\varrho a^{3(w+1)} =$ const., provided a
barotropic equation of state, $p=w\varrho c^2$ for constant $w$, is
assumed. (The condition for phantom models, for example, is $w <
-1$).

From equations (\ref{rho})-(\ref{p}) one can easily see that a
pressure singularity $p \to \mp \infty$ occurs when the acceleration
$\ddot{a} \to \pm \infty$, no matter that the value of the
energy density $\varrho$ and
the scale factor $a(t)$ are regular. Since in that case
$\mid p \mid > \varrho$,
it is clear that the dominant energy condition is violated.
This condition can be achieved if the
scale factor takes the form \cite{barrow04}
\be \label{sf2} a(t) = a_s \left[1 + \left(1 - \delta \right) y^m -
\delta \left( 1 - y \right)^n \right]~, \hspace{0.5cm} y \equiv \frac{t}{t_s} \ee
with the appropriate choice of the constants $\delta, t_s, a_s, m,
n$. Moreover, we can see that the $r$-th derivative of the scale
factor (\ref{sf2}) is given by \bea \label{dotageneral} a^{(r)} &=&
a_s \left[ \frac{m(m-1)...(m-r+1)}{t_s^r} \left(1 - \delta \right)
y^{m-r} \right. \nonumber \\
&+& \left. (-1)^{r-1} \delta \frac{n(n-1)...(n-r+1)}{t_s^r} \left( 1
- y \right)^{n-r}\right]~, \eea and is related to the appropriate
pressure derivative $p^{(r-2)}$. Thus, in general, it is possible
that one has a pressure derivative $p^{(r-2)}$ singularity which
accompanies the blow-up of the r-th derivative of the scale factor
$a^{(r)}$. Observationally this could be manifested in, for example,
the blow-up of the characteristics known as {\em statefinders\/},
such as jerk, snap etc. \cite{plb05}. The pressure derivative
singularity $p^{(r-2)}$ appears when \be \label{r} r-1 < n < r \hspace{1.cm} r
= {\rm integer}~, \ee and for any $r \geq 3$ it fulfills all energy
conditions. These singularities are called generalized sudden future
singularities (GSFS) and are possible, for example, in theories with
higher-order curvature quantum corrections \cite{nojiri}.

Let us now return to the case of $r=2$, for which $1<n<2$ and we
obtain sudden future singularity models of pressure (and obviously
all of its higher derivatives) which lead to violation of the
dominant energy condition. In such models, expressed in terms of the
scale factor (\ref{sf2}), the evolution begins with the standard BB
singularity at $t=0$ for $a=0$, and finishes at SFS for $t=t_s$
where $a=a_s\equiv a(t_s)$ is a constant. (Note that we have changed
the original parametrization of Ref. \cite{barrow04} for the scale
factor (\ref{sf2}) using $A= \delta a_s$).

The standard Friedmann limit (i.e. models without an SFS) of
(\ref{sf2}) is achieved when $\delta \to 0$; hence $\delta$ becomes
the ``non-standardicity" parameter of SFS models. Additionally,
notwithstanding Ref. \cite{barrow04} and in agreement with the field
equations (\ref{rho})-({\ref{p}), we assume that $\delta$ can be
both positive and negative leading to a deceleration or an
acceleration (cf. (\ref{dotageneral})) of the universe,
respectively.

It is important to our discussion that the asymptotic behaviour of
the scale factor (\ref{sf2}) close to the BB singularity at $t=0$ is
given by a simple power-law $a_{\rm BB} = y^m$, simulating the
behaviour of flat $k=0$ barotropic fluid models with $m =
2/[3(w+1)]$~. This allows us to preserve all the standard observed
characteristics of early universe cosmology -- such as the cosmic
microwave background, density perturbations, nucleosynthesis etc. --
provided we choose an appropriate value of $m$. On the other hand,
close to an SFS the asymptotic behaviour of the scale factor is
non-standard, $a_{\rm SFS} = a_s \left[ 1 - \delta \left(1 -
y\right)^n~ \right]$, showing that $a_{SFS}=a_s$ for $t=t_s$ (i.e.
$y=1$) at the SFS. Notice that one does not violate the energy
conditions if the parameter $m$ lies in the range \be \label{m} 0 <
m \leq 1~ \hspace{0.5cm} (w \geq -1/3), \ee
This range of values is, in fact, equivalent to a standard (neither
quintessence-like nor phantom-like) evolution of the universe.
However, with no adverse impact on the field equations
(\ref{rho})-(\ref{p}), one could also extend the values of $m$ to
lie in the complementary ranges \cite{lazkoz} $m > 1$ (i.e.
$-1<w<-1/3$) for quintessence, and  $m < 0$ (i.e. $w<-1$) for
phantom, although these ranges may lead to violation of the strong
and weak energy conditions respectively.

We will next calculate the luminosity distance as a function of
redshift, and hence the redshift-magnitude relation, for SFS models.
This will allow us to establish whether these models are a realistic
possibility for the future evolution of the universe, and more
specifically whether current cosmological observations of high
redshift supernovae are consistent with values of the constant $n$
in the range $1 < n < 2$, as required in order that the scale factor
will display an SFS (or, more generally, a GSFS for $r-1 <n< r$). We
will then explore the range of values for the other SFS model
parameters which are consistent with current observational
constraints on standard cosmology, and thus determine limits on how
far into the future an SFS might occur. In fact, as we will see
below, we need to consider only two further parameters: $\delta$ and
$y_0 = t_0/t_S$, where $t_0$ is the current age of the Universe in
the SFS model. Notice that, in view of (\ref{m}), it is reasonable
to take $m=2/3$ as for the standard dust-dominated evolution.  This
implies that, at early times, our SFS model reduces to the
Einstein-de-Sitter universe.

We proceed within the framework of Friedmann cosmology, and consider
an observer located at $r=0$ at coordinate time $t=t_0$. The
observer receives a light ray emitted at $r=r_1$ at coordinate time
$t=t_1$. We then have a standard null geodesic equation
\be
\label{geod}
\int_0^{r_1} \frac{dr}{\sqrt{1-kr^2}} = \int_{t_1}^{t_0}
\frac{cdt}{a(t)}~,
\ee
with the scale factor $a(t)$ given by (\ref{sf2}). Using (\ref{sf2})
again, the redshift is given by
\be \label{redshift} 1+z=\frac{a(t_0)}{a(t_1)} = \frac{\delta +
\left(1 - \delta \right) y_0^m - \delta \left( 1 - y_0 \right)^n}
{\delta + \left(1 - \delta \right) y_1^m - \delta \left( 1 - y_1
\right)^n}~, \ee
where $y_0 = y(t_0)$ and $y_1 = y(t_1)$. The luminosity distance is
defined as
\bea \label{lumdist} D_L = r_1 a(t_0) \left(1+z \right)~. \eea
Neglecting extinction and $k-$corrections, the observed and absolute
magnitudes of a source at redshift $z$ and luminosity distance $D_L$
are related by
\be \label{m(z)} m(z)=M - 5\log_{10}H_0 + 25 + 5\log_{10} D_L(z),
\ee
which, with the help of the equation (\ref{geod}), (\ref{redshift})
and (\ref{lumdist}), allows a redshift-magnitude relation for SFS
cosmological models to be constructed. It is obvious that equation
(\ref{geod}) has to be integrated numerically in order to establish
the relation between $t_0$ and $t_1$, which can then be inserted
into (\ref{redshift}) and (\ref{lumdist}) to constrain the SFS model
parameters. As a first step we determine the dependence on the SFS
model parameters of the Hubble law, which replaces equation
(\ref{m(z)}) for $z \approx 0$, i.e. $cz \approx H_0 D_L$, where
\bea \label{H0units}
&&H_0({\rm kms}^{-1}{\rm Mpc}^{-1}) = \left(\frac{\dot{a}}{a} \right)_0\\
&=& \frac{3.09 \times 10^{19}}{t_0({\rm sec})y_0} \times \left[
\frac{m \left(1 - \delta\right) y_0^{m-1} + n \delta \left( 1 - y_0
\right)^{n-1}}{\delta + \left(1 - \delta \right) y_0^m - \delta
\left( 1 - y_0 \right)^n}\right]~ \nonumber \eea
is the present value of the Hubble parameter, which we can take as
$72 {\rm km s}^{-1} {\rm Mpc}^{-1}$ \cite{supernovaenew}.

Similarly we could derive an expression, in terms of the SFS model
parameters, for the deceleration parameter $q_0 = -
(\ddot{a}a/\dot{a}^2)_0$. However, in order to search the parameter
space for models which are admissible by current observations, we
write the product of $H_0$ and $q_0$ as
\bea
&&q_0 H_0 = - \left(\frac{\ddot{a}}{\dot{a}} \right)_0 = \\
&-&
\frac{t_0}{y_0} \frac{m(m-1)(1-\delta)y_0^{m-2}
- \delta n(n-1)\left(1 - y_0
\right)^{n-2}}{m(1-\delta)y_0^{m-1} +
n\delta \left(1 - y_0 \right)^{n-1}}~.\nonumber
\eea
In order to obtain an accelerated universe at the present moment of
the evolution, this product should be negative. Fig. \ref{fig1}
shows an example plot of the product $H_0q_0$ as a function of
$\delta$ and $y_0$, with the other parameters fixed at $m=2/3$,
$n=1.9993$, $t_0=13.2457$ Gyr. From the plot we see that there are
large regions of the parameter space which admit cosmic
acceleration. We have explored the parameter space further with
various configurations of $m, n, \delta, y_0, t_0, q_0$, and $H_0$,
and obtained the general conclusion that there is a large class of
SFS models which are compatible with current acceleration.

\begin{figure}[h]
\unitlength1cm
\begin{center}
\scalebox{0.6}{\includegraphics[angle=0]{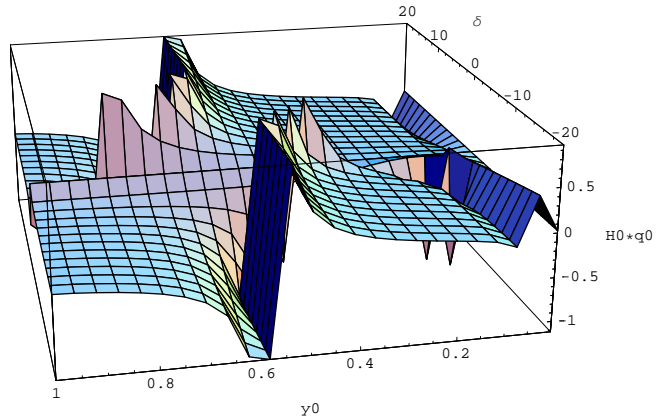}}
\caption{Parameter space $(H_0 q_0, \delta, y_0)$ for fixed values
of $m=2/3, n =1.9993, t_0 =13.3547\,$Gyr of the sudden future
singularity models. There are large regions of the parameter space
which admit cosmic acceleration.} \label{fig1}
\end{center}
\end{figure}

Out of these admissible models we then searched for those which are
compatible with the redshift-magnitude relation (\ref{m(z)})
observed for recent SNIa data \cite{superrecent}, and hence with the
derived parameters of the standard `Concordance cosmology' (CC). We
were able to identify SFS models that are in remarkably tight
agreement with current SNIa data. As an illustrative example Fig.
\ref{fig2} shows luminosity distance as a function of redshift for
the CC model with $H_0 = 72$kms$^{-1}$Mpc$^{-1}$, $\Omega_{m0} =
0.26$ and $\Omega_{\Lambda 0} = 0.74$, and an SFS model with
parameters $m=2/3$, $y_0 = 0.99936$, $\delta = -0.471$, $n=1.9999$.
We see that the SFS model mimics the CC model very closely over a
wide range of redshifts. In particular, it is clear that recent SNIa
data from the Tonry {\em at al.\/} `Gold' sample
\cite{supernovaenew}  and SNLS sample \cite{superrecent} cannot yet
discriminate between the CC and SFS models.

Taking the current age of the universe in the SFS model to be equal
to the age of the CC model, i.e. $t_0 = 13.6$Gyr, we find that the
time to the sudden singularity is $t_s - t_0 \approx 8.7$Myr, which
is {\it amazingly\/} close to the present epoch. In that context
there is no wonder that these singularities are called ``sudden".
We have also checked that the larger the value of $r$
in (\ref{r}) the later in future a GSFS appears. It means that the
strongest of these singularities which violates the dominant energy condition
(i.e. an SFS) is more likely to become reality.

\begin{figure}[h]
\begin{center}
\scalebox{0.3}{\includegraphics[angle=270]{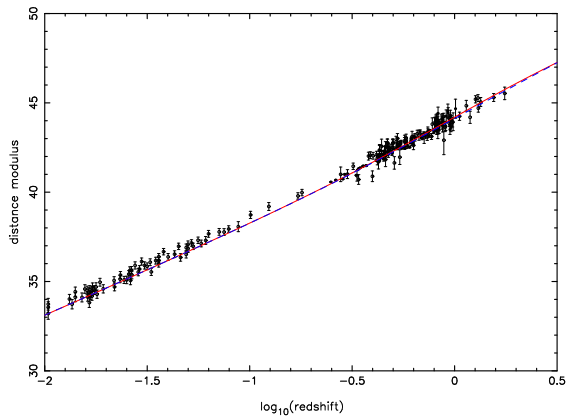}}
\caption{The distance modulus $\mu_L = m - M$ for the concordance
cosmology (CC) model with $H_0 = 72$kms$^{-1}$Mpc$^{-1}$,
$\Omega_{m0} = 0.26$, $\Omega_{\Lambda 0} = 0.74$ (dashed curve) and
sudden future singularity (SFS) model for $m=2/3, n=1.9999, \delta =
-0.471, y_0 = 0.99936$ (solid curve). Also shown are the `Gold'
(open circles) and SNLS (filled circles) SNIa data. Taking the age
of the SFS model to be equal to that of the CC model, i.e.
$t_0=13.6$ Gyr, one finds that an SFS is possible in only 8.7
million years.} \label{fig2}
\end{center}
\end{figure}

Our remark about the effect of the sudden pressure singularity seems
in agreement with the result of Ref. \cite{santos} which showed that
the dominant energy condition is now violated and that it became
violated quite recently (at redshift $z \sim 0.2$). Of course this
violation may also be due to phantom energy \cite{phantom}.

In conclusion, we have shown that a sudden future singularity may
happen in the comparatively near future (e.g. within ten
million years) and its prediction at the present moment of cosmic
evolution cannot be distinguished, with current observational data,
from the prediction given by the standard quintessence scenario of
future evolution in the Concordance Model. Fortunately, sudden
future singularities are characterized by a momentary peak of
infinite tidal forces only; there is no geodesic incompletness which
means that the evolution of the universe may eventually be continued
beyond the SFS until another ``more serious'' singularity such as a
Big-Crunch or a Big-Rip. One could then consider, more generally, a
scale factor of the form \cite{lazkoz,celine} \bea \label{agen} a(t)
&=& A + \left[(a_s - A) - D(t_r - t_s)^p - E t_s^o \right]y^m
\\
&-& \left(A + D t_r^p \right)(1-y)^n + D(t_r - t_s y)^p + E t_s^o
y^o~,\nonumber \eea where the constants $m, o, p, A, D, E$ are chosen so that the
universe begins with a Big-Bang at $t=0$ where $a=0$, next faces a
sudden future singularity at $t=t_s$ where $a(t_s) = a_s$, and then
eventually continues to a Big-Rip at $t=t_r$ where $a(t_r) \to
\infty$. All of the matter sources may be involved since the
constants in (\ref{agen}) can be taken as: $0 < m \leq 1$
(quintessence), $p < 0$ (phantom), and $o>1$ (standard positive
matter pressure).

Whether the universe will end in a Big-Rip or a Big-Crunch is an
open question. Moreover, unlike a sudden future singularity, both a
Big-Rip and Big-Crunch singularity would represent the real end of
the universe.
Fortunately, as was shown in Refs.
\cite{starob00,caldwellPRL}, a Big-Rip singularity is not possible in the
very near future: in order to reach it one must wait about the same
time as the current age of the universe. Apart from that, it is
still possible to avoid it due to a negative tension brane
contribution in a turnaround cyclic cosmology \cite{baum}.

\section{Acknowledgements}

M.P.D. and T.D. acknowledge the support of the Polish Ministry of
Education and Science grant No 1 P03B 043 29 (years 2005-2007).

\end{document}